\documentclass{ijmpd}
\usepackage{graphicx}
\begin{document}
\markboth{D.L.~Wiltshire}
{From time to timescape -- Einstein's unfinished revolution}
%-----------------------------------------------------------------
% Local definitions
\def\PRL#1{Phys.\ Rev.\ Lett.\ {\bf#1}} \def\PR#1{Phys.\ Rev.\ {\bf#1}}
\def\ApJ#1{Astrophys.\ J.\ {\bf#1}}
\def\CQG#1{Class.\ Quantum Grav.\ {\bf#1}}
\def\GRG#1{Gen.\ Relativ.\ Grav.\ {\bf#1}}
\def\beq{\begin{equation}} \def\eeq{\end{equation}}
\def\bea{\begin{eqnarray}} \def\eea{\end{eqnarray}}
\def\Z#1{_{\lower2pt\hbox{$\scriptstyle#1$}}} \def\w#1{\,\hbox{#1}}
\def\X#1{_{\lower2pt\hbox{$\scriptscriptstyle#1$}}}
\font\sevenrm=cmr7 \def\ns#1{_{\hbox{\sevenrm #1}}}
\def\Ns#1{\Z{\hbox{\sevenrm #1}}} \def\ave#1{\langle{#1}\rangle}
\def\kmsMpc{\w{km}\;\w{sec}^{-1}\w{Mpc}^{-1}} \def\bn{\bar n}
\def\dd{{\rm d}} \def\ds{\dd s} \def\al{\alpha} \def\de{\delta}
\def\et{\eta}\def\th{\theta}\def\ph{\phi}\def\rh{\rho}\def\si{\sigma}
\def\ta{\tau} \def\goesas{\mathop{\sim}\limits} \def\Rav{\ave{\cal R}}
%----------------------------------------------------------------
% Include figure files
\def\figvolexp{\centerline{\scalebox{0.3}{\includegraphics{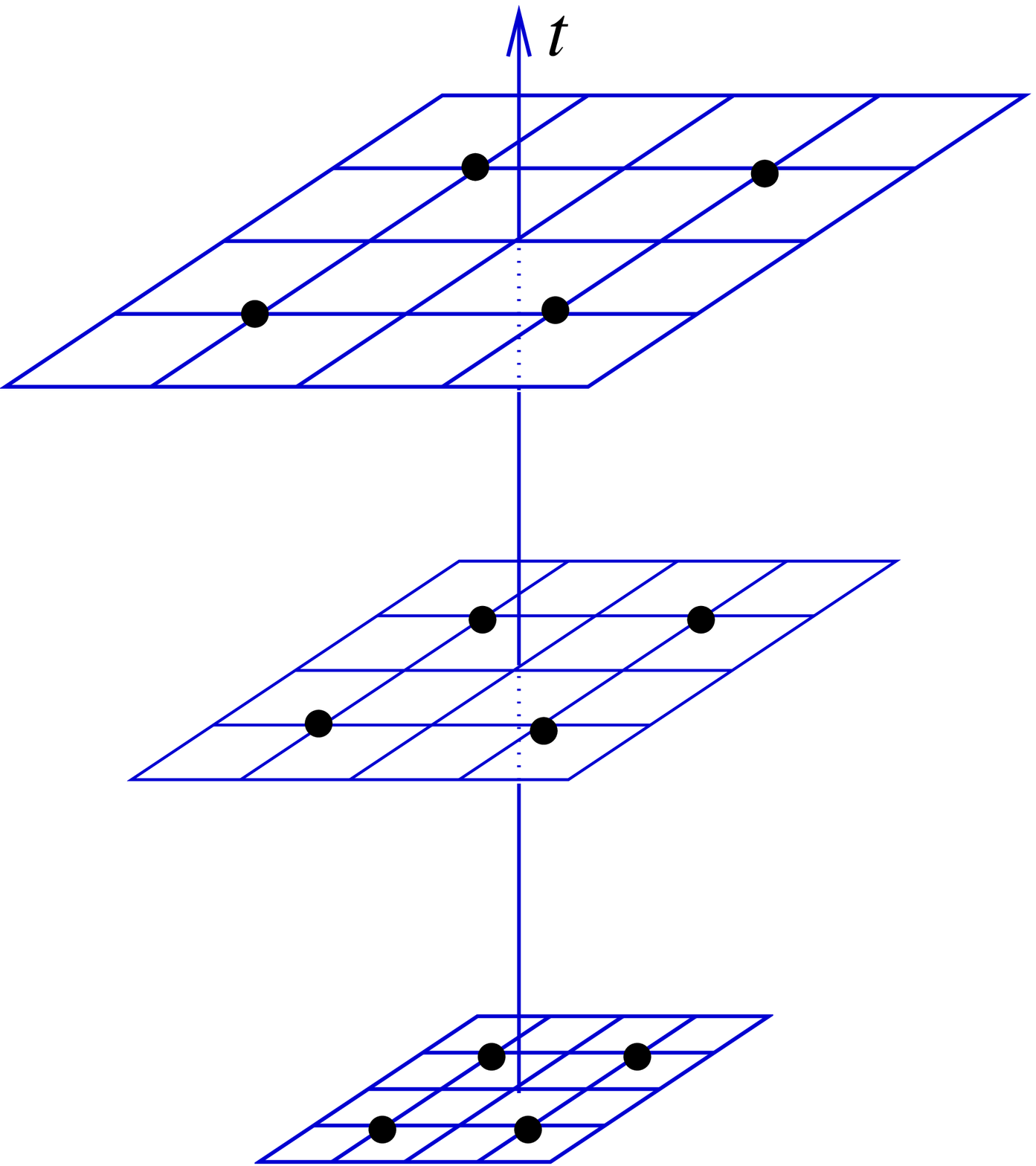}}}}
\def\figlattice{\centerline{\scalebox{0.3}{\includegraphics{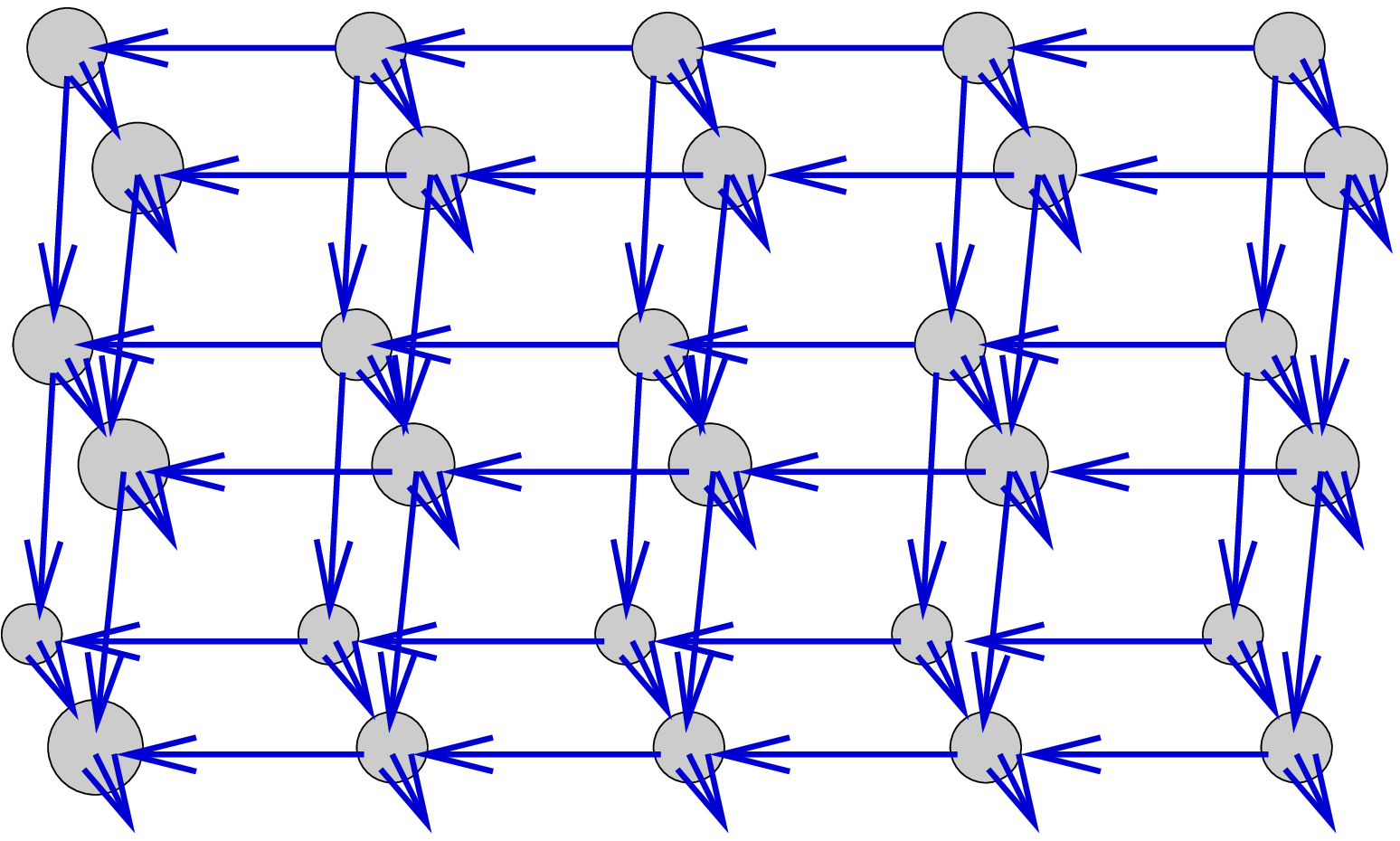}}}}
\def\figequivM{\centerline{\scalebox{0.35}{\includegraphics{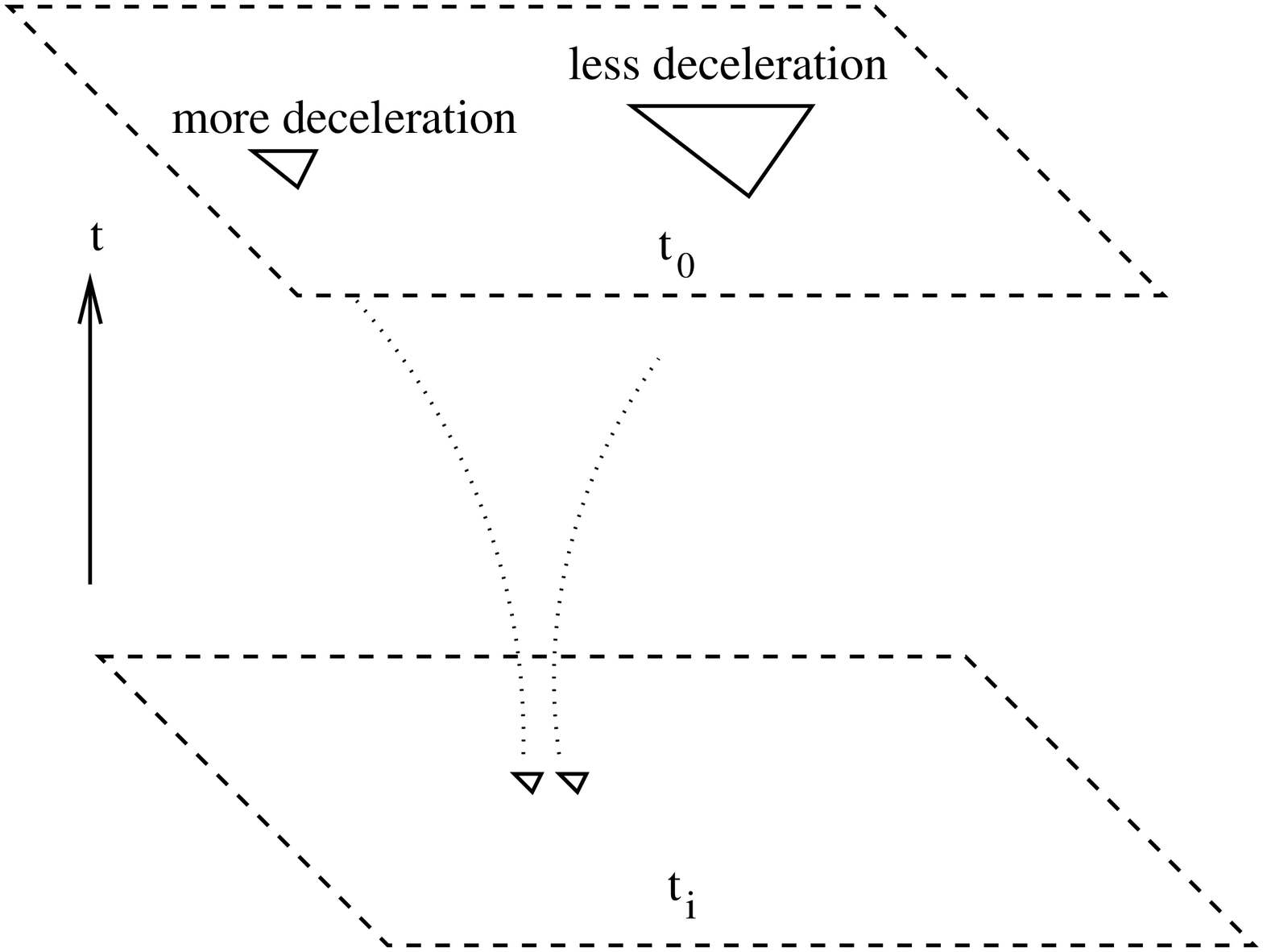}}}
\vskip-20pt\leftline{\bf(a)}\vskip25pt}
\def\figequivGR{\centerline{\scalebox{0.35}{\includegraphics{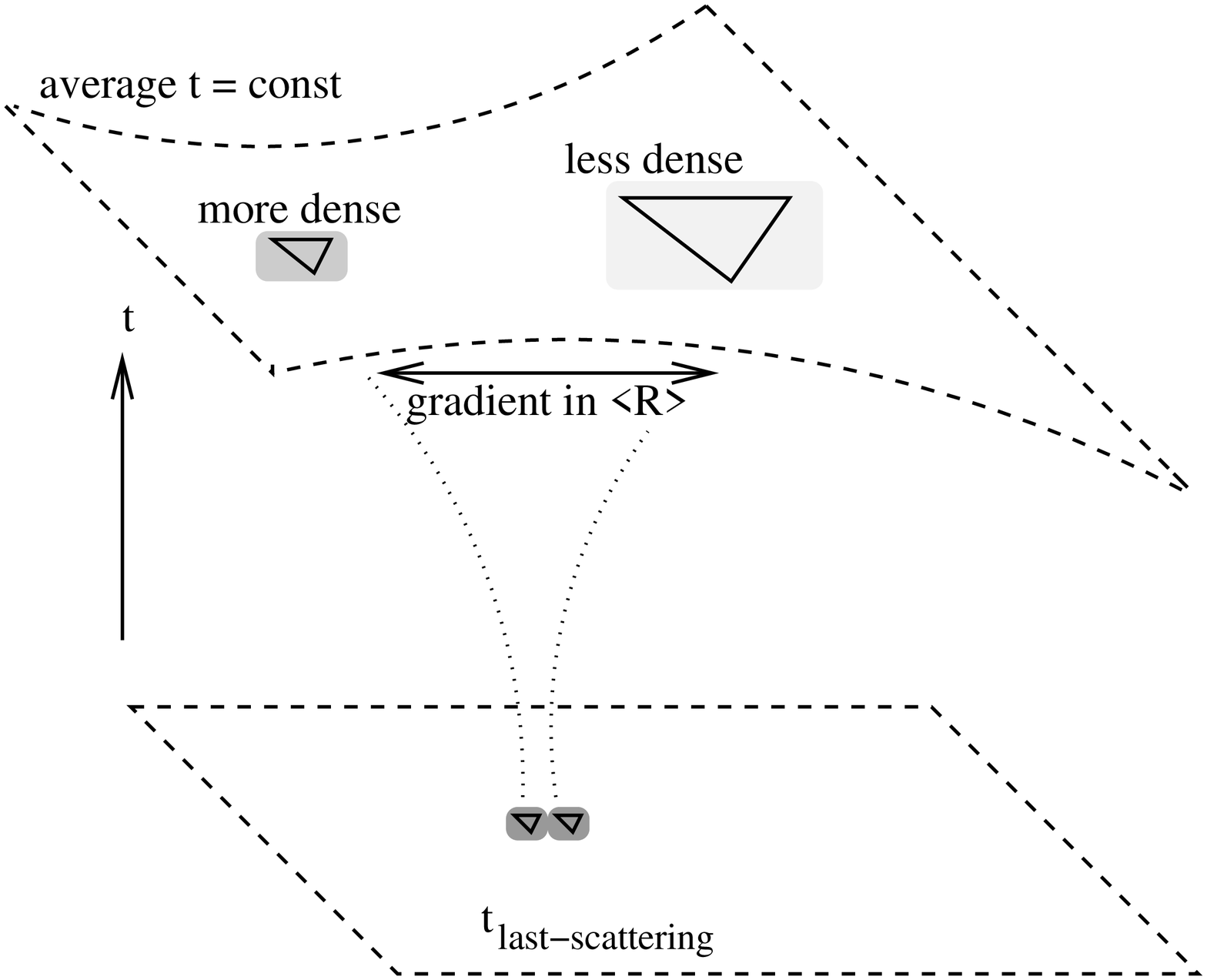}}}
\vskip-20pt\leftline{\bf(b)}\vskip25pt}
%----------------------------------------------------------------
\title{FROM TIME TO TIMESCAPE -- EINSTEIN'S UNFINISHED REVOLUTION\footnote{This
essay was a runner--up in the community awards for the 2008 FQXi Essay Contest
on {\em``the Nature of Time''}.}}
%-----------------------------------------------------------------
\author{DAVID L.~WILTSHIRE}
%-----------------------------------------------------------------
\address{Department of Physics and Astronomy, University of Canterbury,
Private Bag 4800, Christchurch 8140, New Zealand\footnote{Permanent address}\,;
and\\
International Center for Relativistic Astrophysics Network
(ICRANet), P.le della Repubblica 10, Pescara 65121, Italy\\
David.Wiltshire@canterbury.ac.nz}
\maketitle
\def\draftnote{Int.\ J.\ Mod.\ Phys. {\bf D 18} (2009), in press}
%\begin{history}
%\received{Day Month Year}
%\comby{D.V.~Ahluwalia}
%\end{history}

\begin{abstract}
I argue that Einstein overlooked an important aspect of the relativity
of time in never quite realizing his quest to embody Mach's principle in
his theory of gravity. As a step towards that goal, I broaden the Strong
Equivalence Principle to a new principle of physics, the Cosmological
Equivalence Principle, to account for the role of the evolving average regional
density of the universe in the synchronisation of clocks and the relative
calibration of inertial frames. In a universe dominated by voids of the size
observed in large-scale structure surveys, the density contrasts of
expanding regions are strong enough that a relative deceleration of the
background between voids and the environment of galaxies, typically of order
$10^{-10}$ms$^{-2}$, must be accounted for. As a result one finds a universe
whose present age varies by billions of years according to the position of
the observer: a timescape. This model universe is observationally viable: it
passes three critical independent tests, and makes additional predictions.
Dark energy is revealed as a mis-identification of gravitational
energy gradients and the resulting variance in clock rates. Understanding
the biggest mystery in cosmology therefore involves a paradigm shift,
but in an unexpected direction: the conceptual
understanding of time and energy in Einstein's own theory is incomplete.
\end{abstract}
%-----------------------------------------------------------------
\maketitle
%-----------------------------------------------------------------
\section{Introduction}
%-----------------------------------------------------------------

In 1905 Einstein completely changed our understanding of the nature of
time. Rather than being an absolute standard independent of the
physical objects in the universe, time became an intrinsic property of
the clocks carried by the objects themselves. In comparing two clocks,
time could stretch and bend depending on the relative speeds of particles
over their histories.

One hundred years later we find ourselves in a circumstance with
echoes of a century before. Einstein's first revolution, special relativity,
overthrew the then popular aether theories which had been invented to try
to come to grips with the difference between Maxwell's equations for the
propagation of electromagnetic waves on one hand, and Newton's mechanics on
the other. The historical parallels today are striking.
Whereas once the Michaelson--Morley experiment provided evidence
that the Newtonian worldview was flawed, present cosmological observations
suggest that the expansion rate of the universe is accelerating,
posing a foundational problem for our understanding of physics. Again
the first solution that physicists have jumped to is to suppose the
existence of some mysterious fluid, ``dark energy'', which permeates
the fabric of space, the 21st century aether.

In this essay I will argue that just like 100 years ago, the real physics
needed to solve the conundrum of dark energy does not involve a fluid
in the vacuum of space but a deepening of our understanding of the nature
of time, in a manner which many physicists find counter-intuitive. In
particular, time as described by Einstein's second revolution, the general
theory of relativity, is deeply more subtle than the na\"{\i}ve
quasi-Newtonian concept that is applied in the current standard model
of cosmology. The completion of Einstein's second revolution will, I argue,
change our understanding of the universe and the foundations of physics,
by a better understanding of time.

The reason that physicists are quick to invent new forces when confronted
with ``dark energy'', or even to modify gravity in ways that could
change solar system physics, is that we usually think of general relativity
as a completed theory. Yet without even going
to the strong field regime, where the singularity theorems tell us general
relativity does break down, there are deep subtleties in the definition
of energy and momentum in general relativity, which have never been
satisfactorily resolved.

The subtleties, which Einstein and many a mathematical relativist since
have wrestled with, have their origin in the equivalence principle, which
means that we can always get rid of gravity near a
point. As a consequence, the energy, momentum and angular momentum associated
with the gravitational field, which have macroscopic effects on the relative
calibrations of the clocks and rods of observers, cannot be described
by local quantities encoded in a fluidlike energy-momentum tensor.
Instead they are at best {\em quasi-local}.\cite{quasi_rev}

A simple way to understand this is to recall that in the absence of gravity
energy, momentum and angular momentum of objects obey conservation laws. A
conservation law simply means that some quantity is not changing with time.
{\em But whose time?} In general relativity, a dynamical theory of spacetime,
where space and time bend and warp in an evolving manner, a definition of what
is changing or not changing depends on how we split spacetime into spatial
hypersurfaces which evolve with time, and how we choose particular canonical
observers on such surfaces whose clocks are to measure the changes.
Since the mathematical structure of general relativity -- its
diffeomorphism invariance -- does not depend on such choices of observer
frames, there is no unique way to define conservation laws.

The ``quasi-locality'' of gravitational
energy and momentum is very different to a nonlocality of interactions
in flat spacetime which some physicists occasionally postulate and which is
anathema to many, myself included. General relativity is entirely local in
the sense of {\em propagation} of the gravitational interaction, which is
causal. Indeed it thereby overcomes the nonlocality problem of Newtonian
gravity: there is no action at a distance. However, the curved background
on which the interaction propagates may contain its own energy and momentum,
when integrated over sufficiently large regions, and this has
to be understood in the calibration of local rods and clocks at widely
separated events. In dealing with the structure of the whole
universe it is inevitable that we deal with separations on the largest
scales possible.

Since the definition of quasi-local gravitational energy and momentum%
\cite{quasi_rev} depends on spacetime splits that are inherently noncovariant
and nonunique, many questions of naturalness of any particular definition
arise. There is a dilemma that any spacetime split inevitably breaks a
given particle motion into a motion {\em of} the background and a motion
{\em with respect to} the background; and this may involve a degree of
arbitrariness.

The question we are faced with is: which choices of frame have the greatest
utility for the physical description of the universe? I adopt the
view that since quasi-local gravitational energy gradients have their
origin in the equivalence principle, the primary criterion for making
such identifications is that the equivalence principle itself must
be properly formulated, and respected, when making macroscopic cosmological
averages.

\section{Einstein's unfinished principle}

In laying the foundations of general relativity, Einstein sought
to refine our physical understanding of that most central physical concept:
{\em inertia}. As he stated: ``In a consistent theory of relativity
there can be be no inertia relatively to `space', but only an inertia of
masses relatively to one another''.\cite{esu} This is the general
philosophy that underlies Mach's principle, which strongly guided Einstein.
However, the refinement of the understanding of inertia that Einstein
left us with in relation to gravity, the Strong Equivalence Principle, only
goes part-way in addressing Mach's principle.

Einstein's conceptual route began with the {\em Weak Equivalence Principle}
or the {\em Principle of Uniqueness of Free Fall}, known since the
experiments of Galileo, that {\em all bodies subject to no forces other than
gravity will follow the same paths given the same initial positions and
velocities}. Realising that this phenomenological observation implies a
universality for gravity unlike that of other
interactions, Einstein sought to establish gravitation as a property of a
dynamical spacetime structure.
His first step towards that goal was the {\em 1907 Equivalence
Principle}:\cite{eep} {\em All motions in an external static homogeneous
gravitational field are identical to those in no gravitational field if
referred to a uniformly accelerated coordinate system}. In a small sealed
region, an observer on the Earth's surface cannot perform experiments
observationally distinguishable from those in a rocket moving with
acceleration, $\mathbf g$. This is because observers on the Earth's
surface are not inertial observers, but accelerated observers pushed
up by the static forces of the earth beneath our feet. The natural state
is free fall.

The {\em Strong Equivalence Principle} (SEP) then is the statement that even
in an arbitrary gravitational field, by a choice of local coordinates we
can always always find a frame corresponding to the natural state of free fall:
{\em At any event, always and
everywhere, it is possible to choose a local inertial frame (LIF) such that
in a sufficiently small spacetime neighbourhood all non-gravitational laws
of nature take on their familiar forms appropriate to the absence of gravity,
namely the laws of special relativity}. Since we can always eliminate the
effects of gravity near a point, instead of being a force in a
pre-existing space gravity becomes a feature of spacetime structure.
Space and time can curve and bend, and the mathematical object that
describes the bending, the connection, tells us how to relate clocks and rods
of freely falling particles at widely separated events.

This is not the whole story, however, because as yet it tells us nothing
about the spacetime structure of our actual universe. For
that we need to solve Einstein's field equations
\beq G_{\mu\nu}={8\pi G\over c^4}T_{\mu\nu} \label{Einstein}\eeq
to obtain the Einstein curvature tensor, $G_{\mu\nu}$, corresponding to the
distribution of matter sources in the energy-momentum tensor, $T_{\mu\nu}$.
The connection of general relativity
then depends -- via solutions of Einstein's equations -- on the evolving
distribution of matter.

Provided we have solved (\ref{Einstein}) over cosmological scales for
the observed universe, we have addressed Mach's principle which may be stated%
\cite{Bondi,BKL}: {\em ``Local inertial frames are determined through the
distributions of energy and momentum in the universe by some weighted
average of the apparent motions''}. But Einstein never
completed the task of addressing Mach's principle, as he did not specify
what is to be understood by the ``suitable weighted average'' of the
evolving distribution of all the matter fields that can influence
the geometry at any event.

My thesis here is that a further refinement in the understanding
of inertia needs to be made to clarify Mach's principle in
relating local frames to the global universe and to solve
equations (\ref{Einstein}) on cosmological scales. If one views the Einstein
equations as specifying a 4--dimensional continuum completely
determined for all space and all time, if we only knew the distribution
of matter, then the need for further refining the equivalence principle is
easily overlooked. However, general relativity is a causal theory, and the
universe had a beginning. The geometry at any event can only depend on
processes {\it within} its past light cone, limited by the finite age of
the universe. Thus the Einstein equations should be viewed as
dynamical evolution equations for the geometry, limited by
initial conditions with statistical fluctuations.

Einstein overlooked the possibility of further refining the notion of
inertia via the equivalence principle, since the idea that the
universe had a beginning only became widely accepted decades
after he first thought about general relativity. His first journey
through the foundational questions of cosmological relativity had him
worrying about boundary conditions at spatial infinity instead.\cite{esu}
But events at spatial infinity outside the past light cone are
irrelevant if the universe had a beginning.
Although the problem of defining gravitational energy troubled
Einstein greatly, and the relation of the geometry of bound systems to
expanding space was one whose foundational significance was
obvious to him,\cite{ES} once the expanding universe became accepted
he never returned to the equivalence principle
with thought experiments like those he had posed in 1907. I will
take such steps, but first let us recall current standard
practice in cosmology.

\section{Averaging in cosmology\label{average}}

To define a ``suitable weighted average of the apparent motions'' for Mach's
principle requires that we understand the relation between local regional
geometry and average geometry on cosmological scales.\cite{fit}
In solving Einstein's equations for the universe our standard cosmology
still takes the simplifying assumption, made in the first models of Einstein,
Friedmann and Lema\^{\i}tre 80--90 years ago, that the structure of the
universe can be ignored on average, and matter treated as a homogeneous
isotropic fluid. By the evidence of the uniformity of the cosmic
microwave background (CMB) radiation, the universe certainly did satisfy this
approximation when the universe was a few hundred thousand years old and the
first atoms formed. The perturbations in baryons and photons then had an
amplitude $\de\rh/\rh\goesas10^{-5}$ above the mean density, and the
amplitude of perturbations in nonbaryonic dark matter was probably only one
to two orders of magnitude stronger.

At the present epoch, however, following the growth of complex structures from
gravitational collapse, the universe is only statistically homogeneous
if sampled on large scales of order 150--300 Mpc. A box of the size
of statistical homogeneity may be as small as 100$h^{-1}$ Mpc, where $h$ is
the dimensionless parameter related to the Hubble constant by
$H\Z0=100h\kmsMpc$. But within such a box density contrasts $\de\rh/\rh
\goesas-1$ are observed over scales $30h^{-1}$Mpc, which is the typical
diameter of voids which form 40\%--50\% of the volume of the present
universe.\cite{HV1,HV2} If we include the numerous minivoids of smaller
diameters, then the volume of the present universe is dominated by
empty voids, while clusters of galaxies are spread in a cosmic
web of bubble-like sheets that surround the voids, and thin filaments that
thread them.

Over the scales on which $|\de\rh|/\rh\goesas1$ in expanding regions, we
can expect commensurate gradients in Ricci spatial curvature. Our standard
cosmology by contrast assumes a uniform Ricci scalar curvature, and in
applying it we implicitly assume we can ignore spatial curvature gradients and
variations of the relative calibrations of clocks and rods of observers
within cells coarse grained at the scale of statistical homogeneity,
$100h^{-1}$ Mpc. Such an assumption, which effectively assigns one single
cosmic time to the whole universe, has been made for convenience for 80--90
years but is not deeply grounded in theoretical concepts or observational
fact.

One reason that the assumptions of the standard cosmology are not often
questioned, despite the evidence of our telescopes, is that cosmological
gravitational fields are weak due to low average densities of matter. It is
commonly believed that as long as we are in the weak-field limit that we
do not have to worry about the space and time distorting complications of
general relativity, as they only become important near very compact objects
such as neutron stars or black
holes. What is forgotten, however, is that the weak-field limit is always
taken about a background, and once inhomogeneities develop in the universe
there are no exact symmetries to describe the background.

In the absence of an exact symmetries, mathematically described by
Killing vectors, there is no general solution to the problem of
how to keep two clocks synchronized in general relativity. Our usual
intuition about strong and weak gravitational fields is based on
asymptotically flat solutions such as the Schwarzschild and Kerr geometries
which have an exact time symmetry. Since the universe is expanding, however,
no time symmetry exists absolutely. I will argue
that in the absence of such a symmetry a small relative deceleration
of average regional backgrounds can cumulatively lead to large
variations in the clock rates of canonically defined observers.

Numerical simulations of cosmic structure made on large supercomputers
today assume only Newtonian gravity in the background of an expanding
homogeneous universe, whose expansion rate is given by that of a
Friedmann--Lema\^{\i}tre--Robertson--Walker (FLRW) model put in by hand. The
deceleration of the local expansion is not directly coupled to the motion of
the mass particles as it would be in Einstein's equations.

At this point I believe we have overlooked a crucial
foundational question. To make the Newtonian approximation, we must first
make the weak field approximation about a suitable static Minkowski
space. But given that the universe is not static, in choosing an
appropriate Minkowski frame we first have to answer the question: what
is the largest scale on which the SEP can be applied?

\section{The Cosmological Equivalence Principle}

My proposal for applying the equivalence principle on cosmological scales
is to deal with the average effects of the evolving density by extending the
SEP to larger regional frames while removing the time translation and boost
symmetries of the LIF as follows\cite{cep}:

{\em At any event, always and everywhere, it is possible to choose a suitably
defined spacetime neighbourhood, the cosmological inertial frame (CIF), in
which average motions (timelike and null) can be described by geodesics in a
geometry that is Minkowski up to some time-dependent conformal
transformation},
\beq \ds^2\Ns{CIF}=
a^2(\et)\left[-\dd\et^2+\dd r^2+r^2(\dd\th^2+\sin^2\th\,\dd\ph^2)\right].
\label{cif}\eeq

This statement of the Cosmological Equivalence Principle (CEP) reduces to
the standard SEP if $a(\et)$ is constant, or alternatively over very short
time intervals during which the time variation of $a(\et)$ can be neglected.
In those cases the CIF (\ref{cif}) reduces to a LIF. The spatially flat FLRW
metric (\ref{cif}) is to be viewed as a
regional frame, not a geometry for the whole universe.

The SEP says nothing about the average effect of gravity, and
therefore nothing about the ``suitable weighted average of the apparent
motions'' of the matter in the universe. Since gravity for ordinary matter
fields obeying the strong energy condition is universally attractive, the
spacetime geometry of a universe containing matter is not stable, but
is necessarily dynamically evolving. Therefore, accounting for the
average effect of matter to address Mach's principle means that the
relevant frame is one in which time symmetries are removed.

Furthermore,
if we are to demand a smooth Newtonian gravitational limit in all
circumstances, then we have to accommodate the fact that Newtonian gravity
deals with just one scalar source, the density, whereas general relativity
is tensorial. This means that we must be dealing with an average spacetime
with symmetries in taking a Newtonian gravity limit. The metric
(\ref{cif}) removes the time symmetries while preserving the isotropy and
homogeneity of space regionally within a CIF.

%------------------------------------------------------------
\begin{figure}[htb]
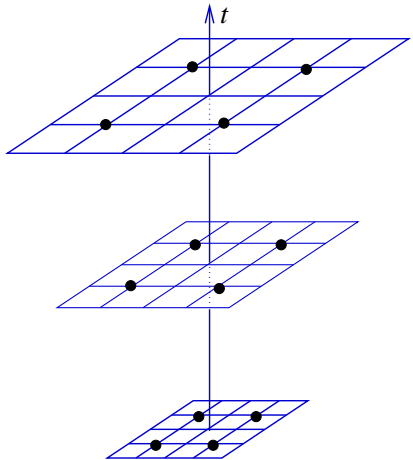

\vbox{\figvolexp
\caption{\label{fig_vol}%
{\sl A set of particles undergoes an isotropic
spatial 3--volume expansion in a spatially flat local region. It is impossible
to locally distinguish the case of particles at rest in a dynamically
expanding cosmological space from particles moving isotropically in a
static Minkowski space. One spatial dimension is suppressed.}}}
\end{figure}
%------------------------------------------------------------

What has this got to do with inertia? Let us first recall the well-known
property that in the case of the volume expanding motions illustrated by
Fig.~\ref{fig_vol}, we cannot {\em locally} distinguish
the case of comoving particles at rest in an expanding metric (\ref{cif})
from the case of particles in motion in the static Minkowski space of the
relevant LIF if we were to choose Riemann normal coordinates.
On local scales, both yield the Hubble law redshift
$$z\simeq {H_0\ell_r\over c},\qquad H_0=\left.\dot a\over a\right|_{t\X0}$$
where $\ell_r$ is the radial proper distance from an observer at the origin
to a source, and an overdot denotes a derivative with respect to $t$, where
$c\,\dd t=a\,\dd\et$. This is true whether the exact relation, $z+1=a\Z0/a$,
is used or the radial Doppler formula $z+1=[(c+v)/(c-v)]^{1/2}$ of special
relativity is used, before making a local approximation.

Rather than simply invoking static special relativistic LIFs over short time
intervals, the CEP demands that we can always find regional frames (\ref{cif})
for arbitrarily long time intervals during which the motion of the particles
is decelerated, $\ddot a < 0$, by the average density of matter. As
Einstein demanded, there should only be inertia of masses relative to
masses. Since the deceleration of the volume expansion is due
to the backreaction of the average density of matter particles in
defining their own background, the CEP thus represents a refinement in the
understanding of inertia. We can always find regional frames (\ref{cif})
in which the average volume-expanding motion with deceleration is such
that {\em we cannot tell whether particles subject to such motion are at
rest in an expanding space, or moving in a static space}. The argument about
whether particles are moving or space is expanding is an argument about
something that is fundamentally indistinguishable.

\section{Thought experiments}

Just as with the original 1907 Einstein equivalence principle, the order
of magnitude of relevant effects can be determined from thought experiments.
To demonstrate this, I will first show that a suitable equivalent
of decelerated Minkowski space particles can always be found for the motion
of a congruence of comoving particles in (\ref{cif}), even for arbitrarily
long time intervals.

\subsection{Semi-tethered lattices}
\begin{figure}[htb]
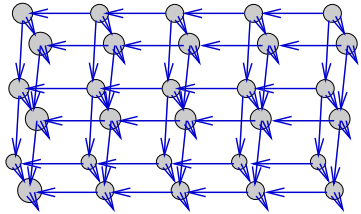

\vbox{\figlattice
\caption{\label{fig_lattice}%
{\sl The semi-tethered lattice. (See text for description.) The time evolution
of the lattice follows a course similar to that of the spatial grid in
Fig.~\ref{fig_vol}, with deceleration.}}}
\end{figure}

Let us construct what I will call the {\em semi-tethered
lattice} by the following means. Take a lattice of Minkowski
observers, initially moving isotropically away from each nearest neighbour at
uniform initial velocities. The lattice of observers
are chosen to be equidistant along mutual oriented $\hat x$, $\hat y$ and
$\hat z$ axes. Now suppose that the observers are each connected to
six others by strings of negligible mass and identical tension along the
mutually oriented spatial
axes, as in Fig.~\ref{fig_lattice}. The strings are not fixed but unwind
freely from spools on which an arbitrarily long supply of string is wound.
The strings initially
unreel at the same uniform rate, representing a ``recession velocity''.
Each observer carries synchronised clocks, and at a prearranged local proper
time all observers apply brakes to each spool,
the braking mechanisms having
been pre-programmed to deliver the same impulse as a function of local
time.
%------------------------------------------------------------
%------------------------------------------------------------

The semi-tethered lattice experiment is directly analogous to
the decelerating volume expansion of (\ref{cif}) due to some average
homogeneous matter density, because it maintains the homogeneity and
isotropy of space over a region as large as the lattice. Work is done in
applying the brakes, and energy can be extracted from this -- just as
kinetic energy of expansion of the universe is converted to other
forms by gravitational collapse. Since brakes are applied in unison, however,
there is {\em no net force on any observer in the lattice}, justifying the
{\em inertial frame} interpretation.
Even if the braking function has an arbitrary time profile, provided it
is applied uniformly at every lattice site
the clocks will remain synchronous in the comoving sense, as all observers
have undergone the same relative deceleration.

\subsection{Relative deceleration of regional backgrounds}
\begin{figure}[htb]
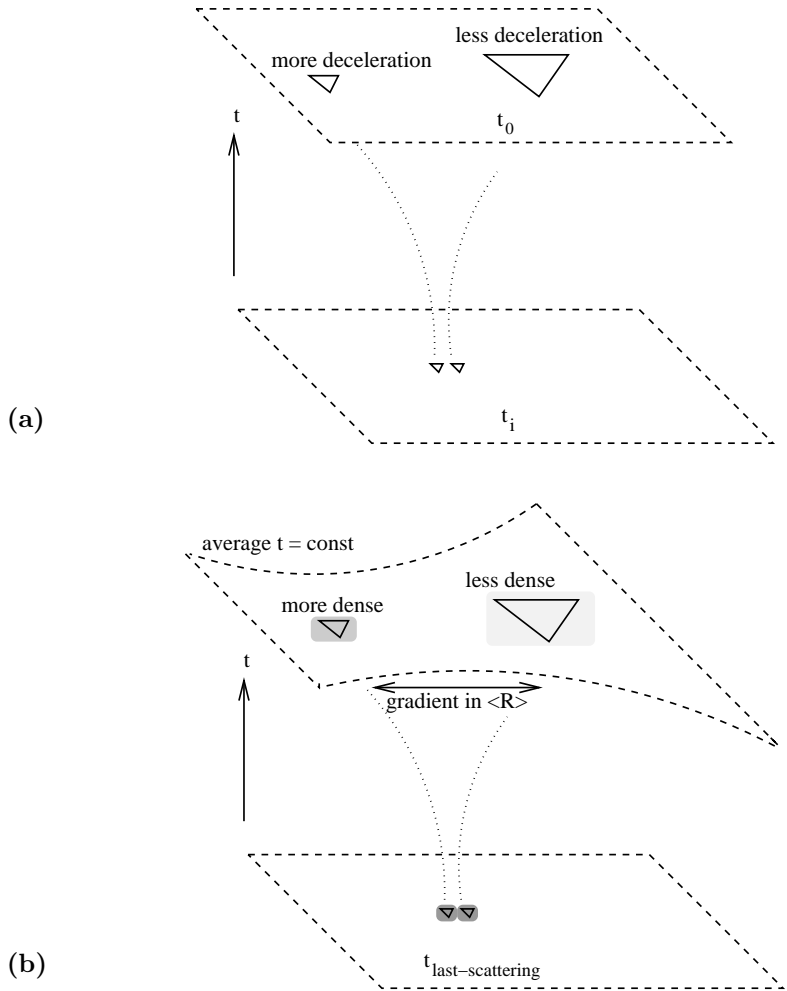

\vbox{\figequivM
\figequivGR
\caption{\label{fig_equiv}%
{\sl Two equivalent situations: {\bf(a)} in Minkowski space observers
in separate semi--tethered lattices, initially expanding at the same
rate, apply brakes homogeneously and isotropically within their respective
regions but at different rates;
{\bf(b)} in the universe which is close to
homogeneous and isotropic at last-scattering comoving observers in separated
regions initially move away from each other isotropically, but experience
different locally homogeneous isotropic decelerations as local density
contrasts grow. In both cases there is a relative deceleration
of the observer congruences and those in the region which has decelerated
more will age less.}}}
\end{figure}

Let us now consider two sets of disjoint semi-tethered lattices,
with identical initial local expansion velocities, in a background
static Minkowski space. (See Fig.~\ref{fig_equiv}(a).)
Observers in the first congruence apply brakes in unison to decelerate
homogeneously and isotropically at one rate. Observers in the second
congruence do so similarly, but at a different rate.
Suppose that when transformed to a global Minkowski frame,
with time $t$,
that at each time step the magnitudes of the 4--decelerations satisfy
$\al\Z1(t)>\al\Z2(t)$ for the respective congruences. By special relativity,
since members of the first congruence decelerate more than those of the
second congruence, at any time $t$ their proper times satisfy $\ta\Z1<\ta\Z2$.
The members of the first congruence age less quickly than members of the
second congruence.
%------------------------------------------------------------
%------------------------------------------------------------

By the CEP, the case of volume expansion of
two disjoint regions of different average density in the actual universe
is entirely analogous. The equivalence of the circumstance rests on the
fact that the expansion of the universe was extremely uniform at the time
of last scattering, by the evidence of the CMB. At that epoch all regions
had almost the {\em same} density -- with tiny fluctuations -- and the same
uniform Hubble flow. At late epochs, suppose that in the frame of any
average cosmological observer there are expanding regions of {\em different}
density which have decelerated by different amounts by a given time, $t$,
according to that observer. Then by the CEP the local proper time of the
comoving observers in the denser region, which has decelerated more,
will be less than that of the equivalent observers in the less
dense region which has decelerated less. (See Fig.~\ref{fig_equiv}(b).)
Consequently the {\em proper time of the observers in the more
dense CIF will be less than that of those in the less dense CIF}, by
equivalence of the two situations.

The fact that a global Minkowski observer does not exist in the second
case does not invalidate the argument. The global Minkowski time is
just a coordinate label. In the cosmological case the only restriction
is that {\em the expansion of both average congruences
must remain homogeneous and isotropic in local regions of different average
density} in the global average $t=$const slice. Provided we patch the regional
frames together suitably, then if regions
in such a slice {\em are still expanding} and have a significant density
contrast we can expect a significant clock rate variance.

This equivalence directly establishes the idea of a {\em gravitational
energy cost for a spatial curvature gradient}, since the existence
of expanding regions of different density within an average $t=$const
slice implies a gradient in the average Ricci scalar curvature, $\Rav$,
on one hand, while the fact that the local proper time varies
on account of the relative deceleration implies a gradient in gravitational
energy on the other.

\section{The timescape and ``dark energy''}

Given the complex structure of voids, walls and filaments described in
Sec.\ \ref{average}, then if we model the universe that we see we must
account for its present epoch inhomogeneity. Buchert's
equations\cite{buch1a,buch1b} provide a suitable framework for describing
the average evolution of Einstein's equations in an inhomogeneous universe,
and give corrections to the Friedmann equations. The interpretation of
Buchert's equations has been controversial.\cite{IW,buch2} The reason for
this stems from the fact that they involve spatial averages.
In general relativity we measure invariants of the local metric, and over
the scales on which the geometry is inhomogeneous the local metric can
vary substantially. Thus every observer cannot be the same average observer.
We must account not only for how inhomogeneity affects average evolution,
but also for how the variance in the geometry affects the calibration of
local clocks and rods relative to the average.

I have developed a new physical interpretation\cite{clocks} of solutions to
the Buchert equations, from the observation that structure formation
provides us with a natural split of scales. We and the objects we observe
are in galaxies which formed from density perturbations that were greater
than critical density, whereas the volume-average location today is in an
underdense void. By the CEP we must account
for the gravitational energy costs of gradients in spatial curvature between
galaxies and the volume-average voids\cite{HV1,HV2} in the
relative calibrations of regional clocks.

The relevant average {\em cosmic rest frame} for the universe is one in which
the underlying regional expansion of CIFs remains uniform in terms of the
rate of change of local proper distances with respect to local proper
times of ideal observers who measure an isotropic CMB.\cite{cep,clocks}
The relation between proper volume and proper diameter is different in
regions of different Ricci curvature. Consequently, even though voids open up
faster when measured by any one set of clocks, since the clocks of isotropic
observers in voids tick faster due to a weaker relative deceleration of their
background, there can still be an underlying uniform local Hubble flow.

There is still a Copernican principle: we are average observers for
observers in a galaxy. However, the local environment of
bound systems which have decoupled from the expansion of space can
differ systematically from the local environment within freely expanding
space in voids. Observers in both locations can measure an isotropic
CMB, but those in voids will measure a cooler mean CMB temperature and
an angular anisotropy scale moved to smaller angles on account of
differences in gravitational energy and spatial curvature respectively.

Cosmic acceleration is an apparent effect\cite{clocks,sol}
which arises when we mistakenly try to fit a Friedmann model to the
whole universe with the incorrect assumption that the local spatial
curvature and local clock rates of isotropic observers everywhere are
identical to our own. An observer in a void will infer no cosmic acceleration,
but observers in galaxies draw different conclusions when converting measured
luminosity distances to an acceleration using two derivatives of a
different time parameter.

The epoch of onset of apparent cosmic acceleration is intimately
tied to the growth of cosmic structure. It begins at a redshift $z\simeq0.9$
when the void fraction reaches 59\%.\cite{clocks,sol}
One finds a model universe,\cite{sol} which by Bayesian comparison
to supernovae data fits at a level statistically indistinguishable from
the standard cosmology with a cosmological constant.$^{17,}$\footnote{{\em
Note added}: An updated analysis of the fit to the most recent supernova data
sets is given in\break Ref.\ \refcite{SW}.}
Furthermore, it matches the angular scale of the sound horizon seen
in the CMB anisotropy spectrum, and the comoving scale of the baryon
acoustic oscillation,\cite{LNW} and may explain other puzzles. Several
tests will enable the model to be distinguished from homogeneous models
with dark energy by future experiments.\cite{obs}

The most startling conclusion is that the age of the universe can vary by
billions of years today depending on whether one is an isotropic observer in a
void or a galaxy. In a galaxy the best-fit age\cite{LNW} is about 14.7 billion
years, at a volume-average position about 18.6 billion years, and in the centre
of a void even larger. This large variance in clocks is counter-intuitive
to physicists because we are talking about weak fields. However, in
the absence of exact symmetries there is no solution to the problem of clock
synchronization in general relativity, even for weak fields. The CEP
extends the conceptual principles of general relativity to address this
problem in a natural manner. Computing the effect\cite{cep} one finds that a
small relative deceleration of backgrounds of differently evolving regional
densities, typically of order $10^{-10}$ms$^{-2}$, cumulatively leads to the
differences claimed when integrated over the lifetime of the universe .

In 1905 Einstein established that time was relative, but in assuming simple
model universes described by a single global frame, and no structure, we
have for the past 80--90 years overlooked the deep possibilities of general
relativity, imagining only a universe described by a single cosmic time.
A universe as inhomogeneous as the one we observe cannot be adequately
described by a single global frame, but if we extend the equivalence
principle to admit regional frames (\ref{cif}), in a manner consistent
with Mach's principle, then the universe that is revealed\cite{clocks,sol}
is a {\em timescape} whose age varies with the inhomogeneous geometry,
a structure much richer in its beauty and subtleties.

In 1917 Einstein realised that {\em in the presence of matter the universe
must change with time}. Faced with the dilemma that this contradicted the
cosmological preconceptions of his time, Einstein introduced a cosmological
constant to try to force the universe to be static.\cite{esu} I believe that
the cosmic mystery of our time, dark energy, requires
that we return to first principles and attempt to think in the way Einstein
taught us to think, rather than compounding his greatest blunder. If I am
correct, then the importance of understanding gravitational energy in
relation to the dynamical nature of time and space is potentially of
foundational importance for quantum gravity too. Einstein's revolution is
not complete.

\medskip {\bf Acknowledgement} This essay condenses some of the arguments of a
longer paper\cite{cep} completed while I was a guest of Prof.\ Remo Ruffini at
ICRANet, Pescara, whom I thank for support and hospitality.
%-----------------------------------------------------------------

\end{document}